\documentclass[article,nofootinbib,nobibnotes]{revtex4}

\usepackage{epsfig}

\def\la{~\mbox{\raisebox{-.6ex}{$\stackrel{<}{\sim}$}}~}

\begin{document}

\begin{flushright}
{\small
{\tt hep-ph/0307379} \\ CITA-2003-32
}
\end{flushright}

\title{Quintessence from Shape Moduli }

\author{Marco Peloso$^1$} \email{peloso@cita.utoronto.ca}
\author{Erich Poppitz$^2$} \email{poppitz@physics.utoronto.ca}

\address{$^1$CITA,
University of Toronto,  Toronto, ON, M5S 3H8, 
Canada\\
$^2$Department of Physics,
University of Toronto, Toronto, ON, M5S 1A7, Canada}

\begin{abstract} 

We show that shape moduli in sub-millimeter extra dimensional
scenarios, addressing the gauge hierarchy problem, can dominate the
energy density of the universe today. In our scenario, the volume of
the extra dimensions is stabilized at a sufficiently high scale to
avoid conflicts with nucleosynthesis and solar-system precision
gravity experiments, while the shape moduli remain light but couple
extremely weakly to brane-localized matter and easily avoid these
bounds. Nonlocal effects in the bulk of the extra dimension generate a
potential for the shape moduli.  The potential has the right form and
order of magnitude to account for the present day cosmic acceleration,
in a way analogous to models of quintessence as a pseudo
Nambu-Goldstone boson.

\end{abstract} 
\date{\today} 

\maketitle 

\section{Introduction and summary}

There is compelling evidence that the universe is undergoing a stage
of accelerated expansion. The original indication from supernovae
data~\cite{sn} is nicely confirmed by a wealth of independent
results~\cite{cmb}. The most immediate explanation for this effect, a
vacuum energy $\rho_\Lambda \sim \left( 0.002 \, {\rm eV} \right)^4
\,$, is more than $120$ orders of magnitude smaller than its naturally
expected value $\sim M_p^4\,$. For this reason, it is often assumed
that---due to some unknown mechanism---$\rho_\Lambda =0\,$, and that
the acceleration is instead due to a new form of energy, named {\it
quintessence}, whose effective equation of state $w$ is sufficiently
close to the one of vacuum. Combining the WMAP measurement of the
cosmic microwave background radiation anisotropies with
supernovae~\cite{sn2} and large scale structure~\cite{2df}
observations, gives the upper bound $w \leq -0.7$ at $95\%$ confidence
level~\cite{cmb}.

As for inflation~\cite{infla}, it is assumed that quintessence can be
effectively described by a scalar field $Q\,$~\cite{quint} which is at
present slowly rolling down some potential $V\,$. One may think that
models of quintessence cannot be much harder to realize than
inflationary ones. After all, primordial inflation lasted for more
than $\sim 60$ e-folds, while the present accelerated stage only
started at redshift $z \sim 0.5$ or so. Hence, the potential of
quintessence does not need to be flat for a region of field values as
large as in the case of the inflaton.  However, the main trouble is
now represented by the much lower scales involved. First, as we noted,
$V \sim \left( 0.002 \, {\rm eV} \right)^4$ is required. Even worse,
the slow roll condition limits the mass of quintessence to be smaller
or comparable to the present Hubble parameter, $m_Q \la H_0 \sim
10^{-33} \, {\rm eV}\,$. Both these requirements make the realization
of particle physics models of quintessence a particularly challenging
task~\cite{kl}.

It is natural to invoke some symmetry to protect such low scales. For
example, supersymmetric models have been
considered~\cite{susyqcd}. However, we know supersymmetry to be broken
at least at the TeV scale. Even assuming that the quintessence sector
only feels this breaking through gravity typically results in too
large a correction to $m_Q\,$. This imposes constraints on models of
quintessence in supergravity~\cite{gravity}. As an alternative
approach, quintessence has been identified with a pseudo
Nambu-Goldstone boson~\cite{pngb}: the quintessence field acquires a
mass due to the explicit breaking of some global symmetry. The reasons
why the breaking could lead to a value of $m_Q$ which is naturally
small are different for the different models considered.

In this work we discuss a different approach, with quintessence arising in
the context of large extra dimensions.  It is sometimes remarked that the
scale $V^{1/4}$ of the present energy density of the universe is very
close to the inverse size of the radii considered in (ADD) models of large
extra dimensions~\cite{add}. Such scenarios are proposed to address the
gauge hierarchy problem; the identification of quintessence with some
``degree of freedom'' of large extra dimensional models would thus relate
in a unique framework the two strongest hierarchies of particle physics.
In the scheme we have in mind, the volume of the extra dimensions is
stabilized at early times, thus guaranteeing standard four dimensional
gravity, while the role of quintessence is played by the moduli fields
controlling the shape of the extra space. The quintessence potential is
due to the Casimir energy of bulk fields. The scale of the potential is
naturally related to the size of the extra dimensions and is protected
from destabilizing corrections by locality and diffeomorphism invariance
of the higher dimensional theory.

An analogous mechanism has been already discussed in the
literature~\cite{qrad}. However, due to the restricted class of shape
deformations considered, it was concluded that it does lead to an
equation of state $w \geq - 1/3 \,$, which is unable to account for
the observed accelerated expansion. We show that a more optimistic
conclusion is reached once generic deformations of the extra space are
taken into account, and that indeed such a mechanism can lead to a
successful model for quintessence. As we will see, the evolution of
the shape moduli strongly resembles the one of quintessence as a
pseudo Nambu-Goldstone boson~\cite{pngb}.~\footnote{Quintessence in
(small) extra dimensions and with a potential similar to the one
of~\cite{pngb} was obtained also in~\cite{qgauge}. In that work, the
extra space is taken to be fixed, and the potential of a Wilson line
of a gauge group in the extra space~\cite{gauge} is employed.}

In this paper, we focus on the case of two large extra dimensions
compactified on a torus, where the generic shape deformations are
known and where the Casimir energy can be accurately computed. 
Our results on the shape moduli as quintessence can be summarized as follows: 
\begin{enumerate}
\item{In a non supersymmetric context, the correct scale for
quintessence is achieved for extra dimensions of size $L \sim 50
\mu$m, corresponding to a fundamental scale of $\sim 7\,$ TeV;
remarkably, such a value of $L$ is expected to be soon tested in short
distance gravity experiments~\cite{boundl}.  However, there is already
a stronger bound (less than one order of magnitude smaller than the
value we are considering) from the necessity to avoid a too rapid
cooling of supernovae~\cite{supernovae}.}
\item{If, instead, the bulk of the extra dimensions is supersymmetric,
while supersymmetry is broken on the Standard Model brane at the
fundamental scale (TeV), the Casimir energy potential of the shape
moduli acquires an additional suppression factor. We show that the
suppression is precisely of the right order of magnitude for the
potential to account for the late-time acceleration, once the scale
$L$ is taken to be consistent with the supernovae bound.}
\end{enumerate}

This paper is organized as follows. In section~\ref{areastab} we discuss
the early time stabilization of the volume of the extra dimensions. In
section~\ref{shapestab} we study the Casimir energy responsible for the
late time evolution of the shape moduli, and their role as quintessence.
The case of a supersymmetric bulk is considered in section~\ref{susystab}.

\section{General setup and area stabilization}
\label{areastab}

In ADD models~\cite{add}, the observed weakness of gravity, i.e. the
high value for the effective four dimensional Planck mass $M_p\,$, is
due to the large size of some extra dimensional space where gravity
propagates. Generally, in models with extra dimensions, $M_p\,$ is
proportional to the volume of the extra space. Variations of $M_p$ are
strongly limited by the requirement of a conventional four dimensional
cosmology from Primordial Nucleosynthesis, starting after about the
first second of the universe, up to the present~\cite{cosmo1,exp}. A
massless radion (i.e. the scalar field whose expectation value
determines the volume of the extra space) also modifies late time
gravity. The radion couples to the trace of the energy momentum tensor
of matter fields. While the coupling is naturally expected to be of
gravitational strength, a further $\sim 10^{-4}$--$10^{-3}$
suppression is needed to fulfill precision tests of general relativity
within the solar system~\cite{gr}. For these reasons, we discuss a
mechanism where the volume of the extra space is stabilized at a
sufficiently high scale.\footnote{An example of the radion as
quintessence is given in \cite{radq}; the specific form of the
potential and kinetic terms and an accurate choice of initial
conditions prevent a conflict of the radion time variation with the
above bounds.}

Stabilization mechanisms for the extra dimensions have drawn
considerable interest. The set-up discussed in~\cite{area} is
particularly simple: two extra dimensions compactified on a Riemann
surface ($S^2, T^2$, etc.) are considered. The area
$\mathcal{A}$ of the surface is stabilized due to an interplay between
the six dimensional cosmological constant and a $U \left( 1 \right)$
field in the bulk. Under the effect of a negative $\Lambda_6\,$, it is
energetically favored for the extra space to shrink. On the other
hand, a gauge field in the extra compact space can have nonvanishing
magnetic flux, which (analogous to the Dirac monopole in four
dimensions~\cite{dirac}) is quantized in units of inverse charge, $
\Phi \equiv \int d X^4 \, d X^5 \epsilon^{\mu \nu} \, F_{\mu \nu} = 2
\, \pi \, N/e$, where $N$ is an integer and $e$ is the gauge coupling
(which has inverse mass dimension). As a consequence, both the
magnetic field and its energy are proportional to the inverse of the
area $\mathcal{A}\,$, contrasting the shrinking of the extra-space due
to the bulk cosmological constant. Minimizing the total energy, one
finds that the area $\mathcal{A}$ is stabilized at:
\begin{eqnarray}
\label{vacuum}
&&\langle \mathcal{A} \rangle = \frac{\mu}{\sqrt{- \, \Lambda_6}}
\quad\quad,\quad \mu \equiv \frac{\pi \, N}{\sqrt{2} \, e} \nonumber\\
\Rightarrow && \frac{M_p^2}{2} \equiv M_6^4 \, \langle \mathcal{A} \rangle
= \frac{M_6^4 \, \mu}{\sqrt{- \, \Lambda_6}} \,\,.
\label{min}
\end{eqnarray}
A large size of the extra dimensions can be achieved with a
sufficiently large number of flux quanta, $N \gg 1\,$. Such a high
value is technically natural, since $N$ can not be changed by
(perturbative) quantum corrections.~\footnote{One might also hope that
a more natural mechanism can be employed, where large extra dimensions
are achieved without exponentially large parameters. Some examples are
provided in the literature, e.g.~\cite{large}.} 

We are interested in studying compactification manifolds which have
shape moduli. The simplest example is a toroidal extra dimension.  If
we place, for example, four branes, with tensions tuned so that each
has deficit angle $\pi$, at the four fixed points of $T^2/Z_2$, we
obtain a space with the topology of a sphere---a ``pillow" with branes
at the four corners---with a metric which has all the moduli of the
torus (given by eqns.~(\ref{metric}, \ref{gamma}) below). We can then
apply the mechanism of \cite{area} to achieve area stabilization.

We should note, however, that the magnetic flux/six dimensional
cosmological constant configuration of \cite{area} is not an exact
solution of the equations of motion, as can be easily checked by
computing the local Einstein equations in the extra dimensions: the
backreaction of the magnetic field on the geometry of the extra space
is neglected in~\cite{area} (since the backreaction can be considered a
perturbation if the energy densities responsible for stabilization are
small in units of $M_6$).  We used the example studied in~\cite{area}
as an illustration, since it only fixes the value of $\mathcal{A}\,$,
leaving the other moduli of the extra space free to evolve; however,
it is not immediately clear to us if this will still be the case once the
backreaction on the geometry is taken into account. 

We take the mechanism of \cite{area} as a strong indication that the
desired stabilization of the area can be achieved. However, this is
not our main focus in this paper; hence, we will not further study
this or other possible area stabilizing mechanisms here.  In the rest
of this work we simply postulate that such a mechanism exists, and
focus on the late time evolution of the shape moduli fields.

\section{Shape stabilization and quintessence}
\label{shapestab}

We concentrate on the case of two extra dimensions compactified on a
torus. This is the simplest example which allows for deformations of
the extra space which preserve a fixed volume. One may regard this
case as a prototype of the generic case in which more extra dimensions
or shape moduli are present.

The torus is parameterized by its area $\mathcal{A}$ and
by two real moduli fields $\tau_1, \tau_2$, defining its shape.  With
the area stabilized, the line interval is given by:
\begin{equation}
\label{metric}
d s^2 = g_{\mu\nu}(x) \, d x^\mu d x^\nu + \gamma_{ij}(x) \, d y^i dy^j \,\,.
\label{line}
\end{equation}
The $x$ coordinates span the noncompact $3+1$ dimensional space,
while $y^i$ are the two extra dimensional coordinates defined in the
interval $y^i \in \left[0 , L \right], L = \langle \mathcal{A}
\rangle^{1/2}\,$.  The four dimensional metric $g_{\mu\nu}$ is the
metric of a Friedmann-Robertson-Walker universe, while the metric
$\gamma_{ij}$ on the torus is:
\begin{equation}
\label{gamma}
\gamma_{ij} = \frac{1}{\tau_2}
\left( \begin{array}{cc} 1 & \tau_1 \\
\tau_1 & |\tau|^{2} \end{array} \right) \,\,,
\end{equation}
where $\tau = \tau_1 + i \, \tau_2\,$, so that $\gamma$ has
determinant one. Integrating over the extra space, we obtain the
following action, describing gravitational and moduli physics at
scales larger than $L$:
\begin{eqnarray}
S_g &=& M_6^4 \int d^6 x \sqrt{-G} R \left( G \right) = \nonumber\\
&=& \frac{M_p^2}{2} \int d^4 x \sqrt{-g} \left[ R \left( g \right) +
\frac{g^{\mu\nu} \partial_\mu \tau \, \partial_\nu {\bar \tau}}{2 \,
\tau_2^2} \right] \,\,,
\label{a4d}
\end{eqnarray}
where $M_p\,$, the (reduced) Planck mass in four dimensions, was given
in eqn.~(\ref{min}), and $\bar\tau = \tau_1 - i \tau_2$.

The Standard Model fields are confined on a $3+1$ dimensional brane
and do not couple to the shape\footnote{Nonlocal effects in the bulk
induce nonderivative couplings between shape moduli and brane-localized
matter. However, in addition to the $1/M_p$ suppression of the radion
coupling to brane matter, the shape moduli coupling is further
suppressed by factors of at least $1/(L M_6) \sim 10^{-15}$.} moduli
$\tau$; they only probe the geometry of the compact space through the
strength of the gravitational interactions set by the volume.  On the
contrary, the Kaluza-Klein spectrum of any bulk field depends on both
the volume and shape of the extra space. Thus, mass differences among
KK eigenstates found in accelerator experiments would allow a
reconstruction of the geometry of the extra space~\cite{shape}.
 
What is more important for us is that the dependence of the KK
spectrum on the shape moduli may provide a stabilization mechanism for
$\tau_{1,2}\,$.  The Casimir energy of a real massless scalar field
obeying periodic boundary conditions in $y^1, y^2$ of
eqn.~(\ref{metric}), generates a potential for the shape moduli given
by:
\begin{eqnarray}
\label{potential}
V_s &=& \frac{1}{L^4} \left[ - \, \frac{4 \, \pi^3 \, \tau_2^3}{945} -
\frac{1}{2 \, \pi^2 \, \tau_2^2} \, \left( \tau_2^2
\frac{\partial^2}{\partial \tau_2^2} - 3 \, \tau_2 \,
\frac{\partial}{\partial \tau_2} + 3 \right) \sum_{p=1}^{\infty} S_p
\right] \,\,, \nonumber\\ S_p &\equiv& \frac{1}{p^5} \, \frac{{\rm
sinh } \left( 2 \, \pi \, p \, \tau_2 \right)}{{\rm cosh } \left( 2 \,
\pi \, p \, \tau_2 \right) - {\rm cos } \left( 2 \, \pi \, p \, \tau_1
\right)} \,\,,
\end{eqnarray}
where we correct an overall $1/2$ factor in eqn.~(61) of \cite{stab}.

While $V_s$ is the Casimir energy computed in a background
(\ref{metric}), with $g_{\mu\nu}$---the Minkowski metric, it is easy
to see that corrections due to the time-dependent nature of $g_{\mu
\nu}$ are of order $H L$, where $H$ is the four dimensional Hubble
scale. Using (\ref{potential}) in a time-dependent background is then
self-consistent provided $H \ll L^{-1}\,$, which is always true for
energies below the fundamental scale of the theory. We also require
that $V_s$ can be considered to be a perturbation of the volume
stabilization mechanism, i.e. the Casimir energy density
(\ref{potential}) should be (at least) smaller than $M_6^4$. The
exponential dependence of~(\ref{potential}) on the canonically
normalized $\phi_2 = \left( M_p / \sqrt{2} \right) \ln \tau_2\,$ thus
limits $\phi_2/M_p < {\cal{O}}(10)$ (moreover, assuming an equally
spaced probability for the initial value of $\phi_2\,$, values $\tau_2
\sim 1$ are more probable than higher values).

It is also important to note that the Casimir potential for the moduli
is a nonlocal effect in the bulk: there exist no local generally
covariant expressions in 6d that generate a potential for $\tau$.
Thus eqn.~(\ref{potential}) is not subject to  divergent radiative
corrections---all divergences are absorbed by local counterterms and
renormalize the coefficients of  various terms in the local action
of the model. Nonlocal quantum gravity effects are also expected to be
exponentially suppressed if the size of the extra dimension is larger
than the fundamental gravity scale $M_6$. The scale of the potential
(\ref{potential}) is thus naturally related to the size of the extra
dimensions.

The contributions of other massless bulk fields are easily related to
(\ref{potential}). For example, the contribution of bulk gravitons and
massless periodic bulk Weyl fermions to the moduli potential is $V_g =
9 V_s$ and $V_\psi= - 4 V_s\,$, respectively \cite{stab}. In the
following, we consider a simple extension of the Standard Model
characterized by the presence of three right handed neutrinos $N$, as
well as gravity. Since $N$ are uncharged under the Standard Model
interactions they can be naturally thought as bulk fields
\cite{Dienes:1998sb}. Moreover, we assume that they are sufficiently
light and so contribute to the Casimir energy as effectively massless
fields (one can also simply postulate the existence of a number of
massless bulk fermions). In this example, the total Casimir energy is
given by:
\begin{equation}
\label{potential2}
V =  9 \, V_s - 3 \times 4 \, V_s = - 3 \, V_s \,\,.
\label{casimir}
\end{equation}

The shape moduli potentials (\ref{potential}, \ref{potential2}),
induced by the Casimir energy, inherit important symmetries coming
from the large diffeomorphisms of the extra compact space, which
relate equivalent tori.  For example, it is obvious from the form of
$S_p$ that $V_s$ is a periodic function of $\tau_1$ with period
one. More generally, equivalent tori are connected by SL$\left( 2 ,\,
Z \right)$ transformations, $\tau \rightarrow \left( a \, \tau + b
\right) / \left( c \, \tau + d \right) \,$, where $a ,\, b ,\, c \,,$
and $d$ are integers obeying $a \, d - b \, c = 1\,$. While this is
less obvious from the explicit expression (\ref{potential}), the
Casimir potential is invariant under the entire group of SL$\left( 2,Z
\right)$ transformations; this fact considerably simplifies its study,
as we see below.

Distinct tori have modular parameters $\tau$ taking values in the
fundamental region $\vert \tau \vert \geq 1 ,\, -1/2 \leq \tau_1 < 1/2
,\, \tau_2 > 0 \,$, so it is enough to study the behavior of the
Casimir energy in this region.  Two values, $\tau_A = i$ and $\tau_B =
{\rm e}^{2 \, \pi \, i/3}\,$, are of special interest, since they are
fixed points of the transformations $\tau \rightarrow -1/\tau$ and
$\tau \rightarrow - 1 / \left( 1 + \tau \right) \,$, respectively (the
${\cal{S}}$ and ${\cal{S}} {\cal{T}}^{-1}$ generators of SL$\left( 2
,\, Z \right)$). Hence, they correspond to extrema of the Casimir
energy~(\ref{potential}). To study their stability, we expand the
potential (\ref{potential2}) up to quadratic order (in practice, only
the first few terms in the sum over $p$ in (\ref{potential}) are
important):
\begin{eqnarray}
L^4 V\vert_{\tau \simeq \tau_A} &\simeq& 0.9 - 0.7 \, \tau_1^2 + 3.4
\left( \tau_2 - 1 \right)^2 \,\,, \nonumber\\
L^4 V\vert_{\tau \simeq \tau_B}&\simeq& 0.8 + 1.6 \left( \tau_1 + 1/2
\right)^2 + 1.6 \left( \tau_2 - \sqrt{3}/{2} \right)^2 \,\,.
\label{exp}
\end{eqnarray}
Thus, $\tau_A$ is a a saddle point of the potential, while $\tau_B$ is
a minimum. In the following, we redefine $V \left( \tau \right)
\rightarrow V \left( \tau \right) - V \left( \tau_B \right) \,$ such
that the minimum is at zero energy; the additional constant
contribution is provided by the mechanism responsible for the
cancellation of the total cosmological constant in the vacuum of the
theory, as it is usually assumed in models of quintessence.
\begin{figure}[h]
\centerline{\epsfig{file=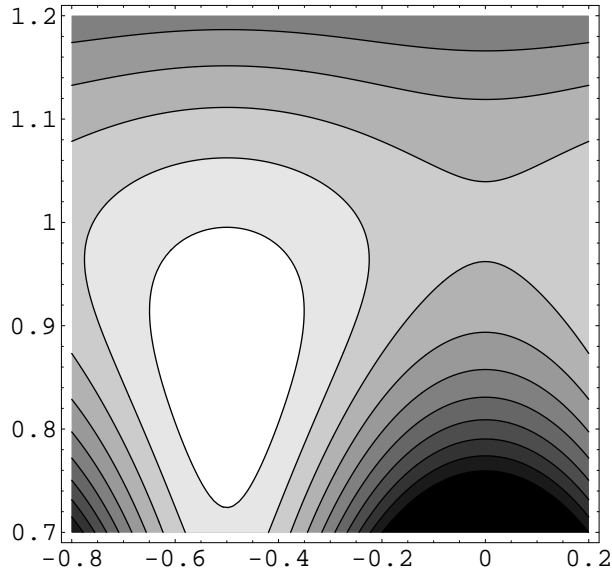,width=0.45 \textwidth}}
\caption{Casimir potential, eqn.~(\ref{potential2}), in the
$\tau_1$-$\tau_2$ plane. The potential is periodic in the $\tau_1$
direction and rises for $\tau_2 \gg 1$ (and $\tau_2 < 1$; the two
regions are physically equivalent, see text). The global minimum,
$\tau_B = e^{2 \pi i/3}$, is indicated in white.}
\label{fig1}
\end{figure}
In figure~\ref{fig1} we show a contourplot of the resulting potential,
in the range of parameters of our interest. This will help illustrate
the dynamics of the two moduli and their role as quintessence fields.

The situation is analogous to the one encountered in models where the
role of quintessence is played by a pseudo Nambu-Goldstone
boson~\cite{pngb} (axion) $\phi$ with lagrangian:
\begin{equation}
{\cal L} = \frac{1}{2} \left( \partial_\mu \phi \right) - U
\quad,\quad U = U_0 \left( 1 - {\rm cos } \: \frac{2 \, \pi \,
\phi}{f} \right) \,\,.
\label{potax}
\end{equation}
In this case, it is assumed that $\phi$ is frozen at the maximum of
the potential $U$ until very recent times. This is possible due to the
friction provided by the expansion of the universe, as evident from
the equation of motion $\ddot{\phi} + 3 \, H \, \dot{\phi} + d U/d
\phi = 0 \,$: at early times, the contributions to the Hubble
parameter $H$ from matter and radiation are high enough to prevent
$\phi$ from moving. As the universe expands, $H$ drops, and eventually
$\phi$ starts rolling towards the minimum of $U\,$. If $\phi$ is
initially sufficiently close to the saddle point, the first part of
its motion occurs in the slow roll regime, and a finite period of
inflation---sufficient to take into account the late time acceleration
of the universe---can take place.

We emphasize that, for this mechanism to work, both the scales of the
potential $U_0$ and of the field $f$ have to acquire very precise, yet
very different values. On one hand, the potential energy of $\phi$
must dominate the present energy density of the universe, leading to
$U_0 \simeq \left( 0.002 \, {\rm eV} \right)^4\,$. On the other hand,
if we want the field to be in the slow roll regime today, rather than
having settled already to the minimum of $U\,$, its effective mass
squared $d^2 U / d \phi^2 \sim U_0 / f^2$ has to be comparable to the
present value of $H^2\,$ (this is probably the main obstacle for
particle physics models of quintessence). As a consequence, $f$ cannot
be taken much smaller than $M_p\,$, unless one is willing to strongly
fine-tune the initial value of $\phi$ unnaturally close to the maximum
of the potential $U\,$. It is rather remarkable that, in the proximity
of the saddle point $\tau_A\,$, the potential~(\ref{casimir}) for the
moduli fields is exactly of the form~(\ref{potax}), with the two
scales $U_0$ and $f$ precisely as required to have a workable model of
quintessence.

As clear from the expansion~(\ref{exp}), motion for $\tau \sim \tau_A$
initially occurs along the $\tau_1$ direction, where the potential $V$
has a cos $\tau_1$ dependence. Since the motion of $\tau$ during the
slow roll regime is rather limited, the potential experienced during
the accelerated stage is effectively of the form~(\ref{potax}) (one
can verify numerically that even a simple quadratic expansion as
in~(\ref{exp}) is sufficient). Thus, the degree of fine tuning on the
initial conditions is the same as the one encountered in models of
Quintessence as a pseudo Nambu-Goldstone boson field~\cite{pngb}.

The scale $f$ of the moduli fields can be easily obtained by direct
inspection of their equation of motion (notice the nonstandard kinetic
term in eqn.~(\ref{a4d})), and it amounts to $f \simeq 0.8 M_p\,$.
Once shifted to have zero cosmological constant in the minimum, the
potential is of order $V \left( \tau \simeq \tau_A \right) \simeq 0.1
/ L^4 \,$. This is of the order of the present energy density of the
universe for a size of the extra dimensions $L \sim 50 \, \mu $m. In
ADD models, this corresponds to a fundamental scale of $\sim 7 \,$
TeV, reasonably close to the electroweak scale. Quite interestingly,
gravity experiments give at present the bound $L \la 150 \, \mu $m,
and are expected to lower their sensitivity to the value we are
considering in the near future~\cite{boundl}.

Despite the fact that $L \sim 50 \; \mu$m is compatible with present
bounds from gravitational experiments, the production of a large
number of light graviton Kaluza-Klein modes leads to rapid cooling of
supernovae and places an upper limit on the value of the radius, $R
\equiv L/(2 \pi)$. For two extra dimensions the bound is $R \le 0.96
\; \mu$m \cite{supernovae}, while our prediction from late-time
domination is $R \sim 7 \; \mu$m.

Thus, it appears that the supernovae constraint---which was not taken
into account in refs.~\cite{radq}, \cite{qrad}---would rule out the
use of shape moduli of two large extra dimensions as
quintessence.~\footnote{There are also cosmological bounds which are
more stringent than the value we are predicting. However, they can be
more easily evaded than the one from supernovae~\cite{cosmo1,cosmo2}.}
However, as we discuss in the next section, if the bulk of the extra
dimension is supersymmetric, while supersymmetry is broken on the
Standard Model brane at the fundamental scale (TeV), the Casimir
energy potential of the shape moduli acquires an additional
suppression factor. We will see that the suppression has just the
right order of magnitude for the potential to account for the
late-time acceleration, even for $L$ small enough to be consistent
with the supernovae bounds.

To conclude this section, we compare our findings to the ones
of~\cite{qrad}, where late time changes of the shape of the extra
space were discussed as well. Only rectangular $n-$tori were
considered in~\cite{qrad}, and an exponential potential---unable to
provide late tame acceleration---was recovered. In our set-up, the
particular class of shape deformations considered in~\cite{qrad}
amounts in varying the modulus field $\tau_2\,$, with $\tau_1\,$ fixed
to zero. If we do so, we also recover an exponential-like potential,
for the canonically normalized field $\phi_2 \propto \ln \tau_2\,$,
cf. eqn.~(\ref{potential}). We see, however, that the dynamics of the
system does not generally proceed through fixed $\tau_1\,$. As we have
discussed, it is precisely the motion along the $\tau_1$ coordinate
which allows for a mechanism suitable for quintessence.

\section{Shape stabilization with supersymmetry in the bulk}
\label{susystab}

In this section, we discuss the evolution of the shape moduli fields
in a supersymmetric context, where bulk fields form complete
supersymmetric multiplets. If supersymmetry is exact, the
contributions to the Casimir energy from bulk fermions and bosons
cancel.  However, supersymmetry must be broken on our brane, with mass
splittings within supermultiplets $\sim {\cal{O}}(1)$ TeV. The
supersymmetry breakdown is then communicated to fields in the bulk.
We assume that the communication is of gravitational strength, giving
a mass splitting within bulk supermultiplets of order $M < {\rm TeV}^2
/ M_p\, \sim 10^{-3}$ eV; this scaling is consistent with the
assumption of brane-localized supersymmetry breaking, whose effect on
the bulk modes should vanish in the infinite volume limit. While the
precise bulk spectrum will certainly depend on the details of the
model (for example, various coupling factors can occur in the above
relation), we stress that our aim here is not to present a complete
model of how supersymmetry breaking and its communication to the bulk
occur. We are rather interested in the order of magnitude effects on
the shape moduli potential.

The mass splitting within the bulk multiplets leads to a nonvanishing
Casimir potential. For completeness and further use below, we give the
expression for the Casimir potential of a massive bulk scalar field
obeying periodic boundary conditions in $y^1, y^2\,$. The massless
case result, eqn.~(\ref{potential}), is replaced by:
\begin{eqnarray}
V_s &=& - \frac{4
\, \left( M R \right)^3 \, \tau_2^{3/2}}{L^4} ~\sum_{p=1}^{\infty}
\frac{K_3 \left( 2 \, \pi \, p \, M R / \sqrt{\tau_2} \right)}{p^3} 
 - \frac{1}{2\,\pi^2\,L^4\,\tau_2^2} \, \left[ \sum_{p=1}^\infty
\left( \tau_2^2 \frac{\partial^2}{\partial \tau_2^2} - 3 \, \tau_2 \,
\frac{\partial}{\partial \tau_2} + 3 \right) {\tilde S}_p \left( x
\right) \right] \Bigg\vert_{x = M R /\sqrt{\tau_2}} \,\,\,\,,
\nonumber\\
{\tilde S}_p \left( x \right) &\equiv& \sum_{n=-\infty}^{\infty} {\rm
e}^{-2 \, \pi \, p \, \tau_2 \sqrt{n^2 + x^2}} \, \frac{{\rm cos }
\left( 2 \, \pi \, p \, n \, \tau_1 \right)}{p^5} \,\,,
\label{massive}
\end{eqnarray}
where $R = L / \left( 2 \, \pi \right)$ and $K_3$ is a modified Bessel
function. As usual, terms that can be absorbed into local bulk or
brane counterterms are omitted from (\ref{massive}); as discussed in
the previous section, such terms are $\tau$-independent. A periodic
fermion (of the same mass) would contribute an amount equal to $-4
V_s$.

In the case under consideration, $M R \ll 1\,$---recall that $R \le
{\cal{O}}(1) \mu$m, while $M < 10^{-3}$ eV---hence it is convenient to
expand the potential in a power series in $M R$. The leading
contribution is obtained for $M=0\,$ and coincides with the massless
expression~(\ref{potential}). However, the fermionic and bosonic
contributions to the Casimir energy cancel for $M=0\,$ (exact
supersymmetry in the bulk). The first nonvanishing contribution occurs
due to the supermultiplet mass splitting and amounts to:
\begin{eqnarray}
{\tilde V} &=& \frac{ \left( M R \right)^2}{L^4} \, \left[
\frac{\pi^3 \, \tau_2^2}{45} + \sum_{p=1}^{\infty} F_p \right]
\nonumber\\
F \left( \tau_1, \, \tau_2\,, p \right) &=& - \frac{2 \, \pi}{p^2} \,
\frac{1 - {\rm cosh } \left( 2 \, \pi \, p \, \tau_2 \right) \, {\rm
cos } \left( 2 \, \pi \, p \, \tau_1 \right)}{\left[ {\rm cosh }
\left( 2 \, \pi \, p \, \tau_2 \right) - {\rm cos } \left( 2 \, \pi \,
p \, \tau_1 \right) \right]^2} + \frac{1}{\tau_2 \, p^3} \, \frac{{\rm
sinh } \left( 2 \, \pi \, p \, \tau_2 \right)}{{\rm cosh } \left( 2 \,
\pi \, p \, \tau_2 \right) - {\rm cos } \left( 2 \, \pi \, \tau_1
\right)} \,\,.
\label{smallm}
\end{eqnarray}
Eqn.~(\ref{smallm}) shows the contribution of a $1/4$ hypermultiplet
where the scalar has mass $M$ while the fermion is massless. The
contributions of other multiplets with supersymmetry breaking mass
splittings can be obtained from eqn.~(\ref{smallm}), as in the
previous section.

Each term in the small-$MR$ expansion of~(\ref{massive}) has the same
SL$(2,Z)$ symmetry as eqn.~(\ref{potential}). In particular, the two
points $\tau_{A,B}$ are also extrema of the potential
term~(\ref{smallm}). The quadratic expansions around these points are:
\begin{eqnarray}
L^4 {\tilde V} \vert_{\tau \simeq \tau_A} &\simeq& \left( M R
\right)^2 \, \left[ 1.92 - 0.5 \, \tau_1^2 + 2.5 \left( \tau_2 -
1 \right)^2 \right] \,\,,\nonumber\\
L^4 {\tilde V} \vert_{\tau \simeq \tau_B} &\simeq& \left( M R
\right)^2 \, \left[ 1.84 + 1.2 \left( \tau_1 + 1/2 \right)^2 +
1.2 \left( \tau_2 - \sqrt{3}/{2} \right)^2 \right] \,\,.
\label{massexp}
\end{eqnarray}
Thus the point $\tau_A = i$ is a saddle point, while $\tau_B = e^{2
\pi i/3}$ is the absolute minimum of the potential. To obtain shape
moduli stabilization, we then require that the contributions of bulk
hypermultiplets (the supersymmetry analogue of the bulk $N$-fields of
the previous section) dominate the contributions of the graviton
supermultiplet.

The situation is thus completely analogous to the one discussed in the
previous section, with one important difference: the scale of the
potential, after subtracting the value at the minimum, is $(0.1/L^4)
(M R)^2$, and is suppressed by the additional $\left( M R \right)^2$
factor, due to the fact that it vanishes in the supersymmetric
limit. Because of this suppression, the Casimir energy can now be of
the order of the present energy density of the universe for an extra
dimension of size $L$ smaller than the $L = 50 \, \mu m$ we found in
the previous section. For example, we can now fix $L \sim 6 \, \mu$m
at its upper allowed limit from \cite{supernovae}. The Casimir energy
is then of the correct scale for $M \sim 2 \cdot 10^{-3}$ eV; since
$V$ scales as $M^2/L^2$ this is also an upper value for $M$ (we absorb
in $M$ various model-dependent factors of order one, coming from,
e.g., the number of bulk multiplets).  As discussed in the beginning
of this section, for a supersymmetry breaking scale on the brane of
order TeV, communicated gravitationally to the bulk, a supersymmetry
breaking splitting of order $10^{-3}$ eV in the bulk is a natural
value. Thus, we conclude that shape moduli in large ($L \sim
{\cal{O}}(1) \mu$m) extra dimensions with TeV-scale brane-localized
supersymmetry breaking naturally provide a viable candidate for
quintessence.

 \vskip1pc
\noindent
{\it Acknowledgements}~~

\smallskip

We thank Keith Dienes, Tony Gherghetta, Joel Giedt, and Bob Holdom for
useful discussions.


\begin{thebibliography}{99}


\bibitem{sn}
 A.~G.~Riess {\it et al.}, Astron.\ J.\ {\bf 116}, 1009 (1998);
S.~Perlmutter {\it et al.}, Astrophys.\ J.\ {\bf 517}, 565 (1999).

\bibitem{cmb}
D.~N.~Spergel {\it et al.}, arXiv:astro-ph/0302209.

\bibitem{sn2}
A.~G.~Riess {\it et al.}, Astrophys.\ J.\  {\bf 560}, 49 (2001).

\bibitem{2df}
M.~Colless {\it et al.}, arXiv:astro-ph/0106498.

\bibitem{infla}
A.~D.~Linde, {\it Particle Physics and Inflationary Cosmology}
(Harwood, Chur, Switzerland, 1990).

\bibitem{quint}
B.~Ratra and P.~J.~Peebles, Phys.\ Rev.\ D {\bf 37}, 3406 (1988);
R.~R.~Caldwell, R.~Dave and P.~J.~Steinhardt, Phys.\ Rev.\ Lett.\ {\bf
80}, 1582 (1998).

\bibitem{kl}
C.~F.~Kolda and D.~H.~Lyth, Phys.\ Lett.\ B {\bf 458}, 197 (1999).

\bibitem{susyqcd}
P.~Binetruy, Phys.\ Rev.\ D {\bf 60}, 063502 (1999); A.~Masiero,
M.~Pietroni and F.~Rosati, Phys.\ Rev.\ D {\bf 61}, 023504 (2000).

\bibitem{gravity}
P.~Brax and J.~Martin, Phys.\ Lett.\ B {\bf 468}, 40 (1999); P.~Brax
and J.~Martin, Phys.\ Rev.\ D {\bf 61}, 103502 (2000); E.~J.~Copeland,
N.~J.~Nunes and F.~Rosati, Phys.\ Rev.\ D {\bf 62}, 123503 (2000).

\bibitem{pngb}
J.~A.~Frieman, C.~T.~Hill, A.~Stebbins and I.~Waga, Phys.\ Rev.\
Lett.\ {\bf 75}, 2077 (1995); S.~M.~Carroll, Phys.\ Rev.\ Lett.\ {\bf
81}, 3067 (1998); J.~E.~Kim, JHEP {\bf 9905}, 022 (1999); K.~Choi,
Phys.\ Rev.\ D {\bf 62}, 043509 (2000); J.~E.~Kim, JHEP {\bf 0006},
016 (2000); Y.~Nomura, T.~Watari and T.~Yanagida, Phys.\ Lett.\ B {\bf
484}, 103 (2000); C.~T.~Hill and A.~K.~Leibovich, Phys.\ Rev.\ D {\bf
66}, 075010 (2002) J.~E.~Kim and H.~P.~Nilles, Phys.\ Lett.\ B {\bf
553} (2003) 1.

\bibitem{add}
N.~Arkani-Hamed, S.~Dimopoulos and G.~R.~Dvali, Phys.\ Lett.\ B {\bf
429}, 263 (1998); I.~Antoniadis, N.~Arkani-Hamed, S.~Dimopoulos and
G.~R.~Dvali, Phys.\ Lett.\ B {\bf 436}, 257 (1998).

\bibitem{qrad}
M.~Pietroni, Phys.\ Rev.\ D {\bf 67}, 103523 (2003).

\bibitem{qgauge}
L.~Pilo, D.~A.~Rayner and A.~Riotto, arXiv:hep-ph/0302087.

\bibitem{gauge}
N.~Arkani-Hamed, H.~C.~Cheng, P.~Creminelli and L.~Randall, Phys.\
Rev.\ Lett.\ {\bf 90}, 221302 (2003).

\bibitem{boundl}
E.~G.~Adelberger [EOT-WASH Group Collaboration], arXiv:hep-ex/0202008.

\bibitem{supernovae} 
S.~Cullen and M.~Perelstein, Phys.\ Rev.\
Lett.\ {\bf 83}, 268 (1999); for a recent update, see: 
 S.~Hannestad
and G.~G.~Raffelt, Phys.\ Rev.\ D {\bf 67}, 125008 (2003);

\bibitem{cosmo1}
N.~Arkani-Hamed, S.~Dimopoulos and G.~R.~Dvali, Phys.\ Rev.\ D {\bf 
59}, 086004 (1999).

\bibitem{exp} P.~Binetruy, C.~Deffayet and D.~Langlois, Nucl.\ Phys.\
B {\bf 565}, 269 (2000); P.~Kanti, I.~I.~Kogan, K.~A.~Olive and
M.~Pospelov, Phys.\ Lett.\ B {\bf 468} (1999) 31; C.~Csaki,
M.~Graesser, L.~Randall and J.~Terning, Phys.\ Rev.\ D {\bf 62} (2000)
045015.

\bibitem{gr}
C.~M.~Will, Living Rev.\ Rel.\ {\bf 4} (2001) 4.

\bibitem{radq}
A.~Albrecht, C.~P.~Burgess, F.~Ravndal and C.~Skordis,
Phys.\ Rev.\ D {\bf 65}, 123507 (2002).

\bibitem{area}
E.~Cremmer and J.~Scherk, Nucl.\ Phys.\ B {\bf 108}, 409 (1976);
R.~Sundrum, Phys.\ Rev.\ D {\bf 59}, 085010 (1999); N.~Arkani-Hamed,
S.~Dimopoulos and J.~March-Russell, Phys.\ Rev.\ D {\bf 63}, 064020
(2001).

\bibitem{dirac}
P.~A.~Dirac, Phys.\ Rev.\  {\bf 74}, 817 (1948).

\bibitem{large} N.~Arkani-Hamed, L.~J.~Hall, D.~R.~Smith and
N.~Weiner, Phys.\ Rev.\ D {\bf 62}, 105002 (2000); Z.~Chacko,
P.~J.~Fox, A.~E.~Nelson and N.~Weiner, JHEP {\bf 0203}, 001 (2002);
A.~Albrecht, C.~P.~Burgess, F.~Ravndal and C.~Skordis, Phys.\ Rev.\ D
{\bf 65}, 123506 (2002).

\bibitem{shape}
K.~R.~Dienes, Phys.\ Rev.\ Lett.\  {\bf 88}, 011601 (2002).

\bibitem{stab}
E.~Ponton and E.~Poppitz, JHEP {\bf 0106}, 019 (2001).

\bibitem{Dienes:1998sb}
K.~R.~Dienes, E.~Dudas and T.~Gherghetta, Nucl.\ Phys.\ B {\bf 557},
25 (1999)
[arXiv:hep-ph/9811428]; N.~Arkani-Hamed, S.~Dimopoulos,
G.~R.~Dvali and J.~March-Russell, Phys.\ Rev.\ D {\bf 65}, 024032
(2002).

\bibitem{cosmo2}
L.~J.~Hall and D.~R.~Smith, Phys.\ Rev.\ D {\bf 60}, 085008 (1999).




\end{thebibliography}
\end{document}